\input o.sty
\def\dx{\partial _x}\def\dy{\partial _y}\def\dz{\partial _z}\def\dw{\partial
_w}
\def\dm{\CD^1(M)} \def\dmc{\CD^1(M,\C)} \def\dmh{\CD^1(\Mhat)}
\def\vm{\CV(M)} \def\vmc{\CV(M,\C)} \def\vmh{\CV(\Mhat)}
\def\fm{\CF(M)} \def\fmc{\CF(M,\C)} \def\fmr{\CF(M,\R)} \def\fmh{\CF(\Mhat)}
\def\g{\frak g} \def\gc{\g_\C}    \def\gh{\widehat\g} \def\gt{\widetilde\g}
\def\h{\frak h} \def\hc{\h_\C}    \def\hh{\widehat\h}
\def\htt{\widetilde\h} \def\mt{\widetilde\m} \def\hit{H^1(\htt,\qt)} \def\qt{\widetilde\q}
\def\fmt{\CF\bigl(\Mtilde\bigr)}
\def\q{\CQ} \def\qc{\q_\C}   \def\qh{\widehat\q}
\def\m{{\frak M}}
\def\mc{\m_\C}   \def\mh{\widehat\m}
\def\n{\frak N} \def\nc{\n_\C}   \def\nh{\widehat\n}
\def\zz{Z^1(\h,\q)} \def\zh{Z^1(\hh,\qh)} \def\zr{Z^1(\h,\q)} \def\zc{Z^1(\hc,\qc)}
\def\bb{B^1(\h,\q)} \def\bh{B^1(\hh,\qh)} \def\br{B^1(\h,\q)}\def\bc{B^1(\hc,\qc)}
\def\hi{H^1(\h,\q)}  \def\hih{H^1(\hh,\qh)}  \def\hir{H^1(\h,\q)} \def\hic{H^1(\hc,\qc)}
\def\ran#1#2{0\leq  #1 \leq #2}
\def\Case#1{\is{Case 4.#1}:\quad}
\def\vf.{vector field}
\def\fo.{first-order}
\def\do.{differential operator}
\def\fd.{finite-dimensional}
\def\Fd.{Finite-dimensional}
\def\qes.{quasi-exactly solvable}
\def\Qes.{Quasi-exactly solvable}
\def\sc.{Schr\"odinger}
\def\with{\roq{with}}
\def\forall{\roq{for all}}
\let\ph\varphi
\def\Strut{\vphantom{{T^T}^T}}
\catcode`\@=11
\def\Eqalign#1{\null\,\vcenter{\openup\jot\m@th
  \ialign{\strut\hfil$\displaystyle{##}$&
    $\displaystyle{{}##}$\hfil&&
    \qquad\hfil$\displaystyle{##}$&
    $\displaystyle{{}##}$\hfil\crcr
  #1\crcr}}\,}
\catcode`\@=12
\parskip = 2pt plus 2pt minus 2pt
\widetildes
\widehats
\barrs

\proofmodefalse
\refsin
\keyin

\Title
Real Lie Algebras of Differential Operators \\
and Quasi-Exactly Solvable Potentials.

\author
Artemio Gonz\'alez-L\'opez\\
Departamento de F\'\i sica Te\'orica II\\
Universidad Complutense\\
28040 Madrid\\
SPAIN 
\email artemio@ciruelo.fis.ucm.es
\support DGICYT Grant PB92--0197.

\author
Niky Kamran\\
Department of Mathematics\\
McGill University\\
Montr\'eal, Qu\'ebec\\
CANADA \quad H3A 2K6
\email nkamran@math.mcgill.ca
\support an NSERC Grant.

\pjo

\printauthor
\dated
\Abstract We first establish some general results connecting real and
complex Lie algebras of \fo. differential operators. These
are applied to completely classify all finite-dimensional
real Lie algebras of first-order differential operators in 
$\R^2$.  Furthermore, we find all algebras which are quasi-exactly solvable,
along with the associated finite-dimensional modules of analytic functions. 
 The resulting real Lie algebras are used to construct new quasi-exactly 
solvable Schr\"odinger operators on $\R^2$.\par

\page
\vglue .25in
\Section i Introduction.

Over the past decade, a great deal of effort has been directed towards the construction of 
physically significant quantum mechanical systems for which the spectral problem 
associated to the Schr\"odinger operator may not be exactly solvable, but for which at least 
part of the spectrum can be computed exactly by algebraic methods. 
The concept of a ``spectrum generating algebra"  appears to date back to a 
1959 paper of Goshen and Lipkin, \rf{GoshLip},
 which seems to have remained largely unnoticed.
Spectrum-generating algebras were independently rediscovered in 1965
 by two groups of theoretical physicists, \rf{BarutBohm}, 
\rf{DGN}.  This served as the impetus for a great deal  of 
subsequent research activity, as
evidenced by the two volume set of reprints \rf{BNB}, and
the more recent conference proceedings \rf{GO}.  A survey of
the history and contributions appears in the review paper of
Bohm and Ne'eman, \rf{BN}, at the start of these
volumes.  Applications of spectrum generating algebras to
molecular spectroscopy began in the early 1980's with the work of
Iachello, Levine, Alhassid, G\"ursey, and collaborators; see the recent book,
\rf{IL}, for a survey of the theory and applications. In both the nuclear and
spectroscopic applications, the relevant Hamiltonian is a 
Lie-algebraic differential operator, meaning 
that it lies in the universal enveloping algebra of a finite-dimensional Lie algebra of
first-order differential operators --- the ``hidden symmetry algebra" of the
quantum mechanical problem. 

In the late 1980's, Shifman, Turbiner, Ushveridze,
and their collaborators undertook the analysis of a significant new class of
Schr\"odinger operators which they called ``quasi-exactly solvable",
\rf{ShifTurb}, \rf{Turbiner}, \rf{Ush}. A Lie-algebraic differential operator
satisfies the additional condition of quasi-exact solvability provided its
hidden symmetry algebra admits a finite-dimensional representation space (or
module) consisting of smooth functions on the underlying space. Since the
differential operator is a polynomial in the \fo. operators generating
the symmetry algebra, it leaves the associated module invariant, and hence
restricts to a finite-dimensional linear transformation thereon. If the
functions in the module are square-integrable with respect to the natural
measure, the quasi-exactly solvable operator is said to be normalizable; in
this case, the eigenvalues of the restricted linear transformation will
correspond to part of the point spectrum of the differential operator,
thereby reducing the computation of that part of the spectrum to an algebraic
eigenvalue problem. See \rf U for a comprehensive survey of the theory and
applications of quasi-exactly solvable systems in physics.

In non-relativistic quantum mechanics one is interested in 
real-valued \is{Schr\"odinger operators} of 
the form
$ H = -\Delta + V$,
where $\Delta$ 
is the Laplacian (kinetic energy) operator on a 
finite-dimensional real Riemannian manifold\fnote{More generally, 
for multi-particle
systems, the kinetic energy operator would be a sum
of Laplacians.  But such systems fit readily into our general
framework, since a sum of Laplacians is just the Laplacian operator on
a Cartesian product manifold whose metric tensor is 
the direct sum of the individual metrics.} $M$,
and $V$ is the potential. (The physical units are taken so that
$\hbarq=2m=1$.)  Thus, $H$ is quasi-exactly solvable 
if there exists a finite-dimensional Lie algebra $\g$ of
first-order differential operators, admitting a finite-dimensional module
$\n$  of smooth functions, such that $H$ can be 
written
as a bilinear combination
$$ H = \sum_{a,b=1}^r C_{ab}{T^a}{T^b} + \sum_{a=1}^r C_{a}T^{a} \Eq{h2}
$$
of the generators $T^a$ of $\g$. Here $r = \dim\g$, the
coefficients $C_{ab}$ and
$C_{a}$ are  real constants, and we have omitted an irrelevant constant term
that can be absorbed in the  energy.  Since the quantum mechanical wave
functions are, in general, complex-valued, the module $\n$ can contain
complex-valued functions.  Clearly, since each of the generators of $\g$
leaves $\n$ invariant, $T^a(\n)\subset \n$, the Hamiltonian
$H$ admits $\n$ as a finite-dimensional invariant space, $H(\n)\subset \n$.  Thus,
assuming again that the functions in $\n$ are normalizable, the
spectrum of a quasi-exactly solvable Schr\"odinger operator $H$ has an algebraic sector, 
which
can be computed algebraically by diagonalizing the finite-dimensional linear operator 
obtained by restricting $H$ to $\n$.  
In the decomposition \eq{h2}, even though $H$ is required to be a 
real differential operator, the  generators
$T^a$ of the hidden symmetry algebra $\g$ could conceivably be complex-valued, 
with complex coefficients $C_{ab}$, $C_a$ also.  In
practice, however, if the operators
$T_a$ are complex-valued the conditions on the complex constants $C_{ab}$ 
and
$C_a$ arising from the fact that $H$ must be a \is{real} differential operator --- \ie the
potential $V$ must be a real-valued function --- are virtually impossible to satisfy.
For instance, 
extensive symbolic algebra calculations using the realization of $\so{3,\R}$
given by Case 4.2 in Table 4 failed to yield any real-valued potentials, even
after allowing for a possible rescaling of the wave function by a non-zero gauge factor, \cf
\eq{gauge}.
Therefore, the primary objects of interest for the construction of real
\qes. \sc. operators are \fd.  real Lie
algebras of real-valued \fo. \do.s on an open subset
$M$ of $m$-dimensional Euclidean space, which admit a \fd. module of (smooth)
\is{complex-valued} wave functions.  It is the classification of such objects,
particularly in the two-dimensional case, that forms the primary focus of this
paper.

One of the
principal research goals in this subject has been to obtain and classify new
and physically important examples of quasi-exactly solvable Schr\"odinger
operators. In the one-dimensional case, complete results are known. In this
case, the real and complex classifications are identical, since, up to equivalence,
there is essentially just one
family of one-dimensional quasi-exactly solvable Lie algebras of \fo.
differential operators, indexed by a single quantum number $n\in \N$. The
symmetry algebra can be identified with the unimodular Lie algebra $\sLC2$ 
corresponding to the projective group action, having its standard
representation on the space of polynomials of degree at most $n$. The
complete list of one-dimensional quasi-exactly solvable Schr\"odinger
operators was found in \rf{Turbiner}; further, a complete solution to the
normalizability problem for these operators was recently determined, 
\rf{GKOnorm}, \rf{GKO}. The higher-dimensional case is much more challenging, 
owing notably to the fact that, already in the case of planar vector fields, there are infinitely 
many distinct finite-dimensional Lie algebras of vector fields, of arbitrarily large 
dimension.
Moreover, some of the complex Lie algebras have several different inequivalent real forms, 
and so the classification of complex quasi-exactly solvable Lie algebras of
differential  operators does not fully resolve the corresponding real problem.
In the two-dimensional  case, a complete list of Lie algebras of \fo.
differential operators in two complex  variables was found by us in
\rf{GKOCanada}, \rf{GKOlado}. Our starting point was  Lie's complete
classification of the finite-dimensional Lie algebras of vector fields in two 
complex variables, \rf{LieT}, \rf{Liegr}, and then applying methods based on
Lie algebra  cohomology to determine the associated Lie algebras of
differential operators. In 
\rf{GKO2}, this classification was applied to construct several new families of 
normalizable quasi-exactly solvable Schr\"odinger operators in two dimensions, on both 
flat and curved spaces. The complete classification of (normalizable) quasi-exactly solvable 
Schr\"odinger operators remains to be done, although this appears to be an extremely 
difficult
problem. In this paper, we complete the local 
classification of Lie algebras of
\fo. differential operators in two real variables, using as our starting point
the classification of Lie algebras of vector fields in $\R^2$ that was 
rigorously established in \rf{GKOreal}. Interestingly, for the five additional real forms
not appearing in Lie's complex classification, every associated Lie algebra of differential 
operators 
is a subalgebra of a discrete family of algebras isomorphic to the pseudo-orthogonal 
Lie algebra $\so{3,1}$ --- the linear isometries of Minkowski space.
Our classification will be applied to find a few interesting new 
examples of real quasi-exactly solvable Schr\"odinger operators in two-dimensional space. 

As for the extension of these results to more than two dimensions, the difficulty is that 
there is still no complete classification of even the Lie algebras of vector fields in $\C^3$. Lie, \rf{LieT}, describes a substantial fraction
of the classification, and indicates that he had completed it, although, as
far as we know, this has never appeared in print! Nevertheless, it would be
an interesting and physically relevant problem to extend the classification
even to those examples discussed by Lie. Our methods for
passing from a complex Lie algebra of differential operators to its 
associated real forms and the quasi-exact solvability conditions are of
general applicability, and can be readily applied to higher dimensional
situations for which one can determine relevant real forms.

The paper is organized as follows. In \section{la}, we review the basic theory of Lie 
algebras of differential operators, describing their characterization based on a Lie algebra of 
vector fields, an associated module or representation space, and a cohomology class in the 
associated Lie algebra cohomology. In \section{c} we describe the connection 
between complex Lie algebras of differential operators and their real-analytic counterparts. 
The basic result that allows us to straightforwardly pass from real forms of the Lie algebra 
of 
vector fields to corresponding real forms of the Lie algebra of differential operators is that 
any complexified Lie algebra of vector fields with a complexified module automatically has 
complexified cohomology. In \section{qes} we review the basic facts about quasi-exact 
solvability and the quantization of cohomology in the complex case, and its real-analytic 
counterparts. In \section{2d}, we complete the classification of real Lie algebras of 
differential operators in the plane along with the quasi-exact solvability condition and 
associated modules. In particular, we show that, of the five additional Lie algebras of 
vector fields in the real plane not being simple restrictions of the complex ones in Lie's list, 
the only one admitting an extension to a Lie algebra of first-order differential operators by a 
non-trivial real-valued cocycle is a family of central extensions of
$\so{3,1}$, for which the  corresponding Lie algebras of first-order
differential operators are labeled by a non-negative integer. As an
application, we obtain the explicit decomposition into irreducible submodules
of the modules obtained by Morozov {\it et al.}\/, \rf{MPRST}, for the case of
$\so{3,\R}$. Finally, in \section{do}, we review the theory of quasi-exactly
solvable Schr\"odinger operators, and use the preceding results to both
reconstruct the family of operators found by Zaslavskii,
\rf{Zas}, as well as exhibit
 new real quasi-exactly solvable potentials in two variables.


\Section{la} Lie Algebras of Differential Operators.

First, we review the basic theory underlying the classification of Lie algebras of
differential operators in the complex domain, referring the reader to \rf{GKO} for details. 
Let
$M$ be a complex-analytic manifold\fnote
{In our applications, $M$ will always be an open subset of Euclidean space, 
but the basic theory in this section is valid in general.}
 of dimension $m$. Let $\fm = \fmc$ denote the space 
of complex-valued analytic
functions on
$M$. Let $\vm $ denote the space of analytic vector fields on
$M$, which forms an infinite-dimensional Lie algebra based on the standard Lie bracket
operation $[\bov,\bow]$. The space $\fm$ forms a \is{module}, or \is{representation 
space},
for the Lie algebra $\vm $, which acts via derivations: $h\mapsto \bov(h)$ for $\bov\in \vm 
$,
$h\in \fm$. Similarly, $\fm $ acts on itself by multiplication, \ie a function $f \in \fm $
defines the linear multiplication operator $h\mapsto f \cdot h$ for $h\in \fm $. The Lie
algebra of \fo. differential operators on $M$ can be identified with the semidirect
product of these two Lie algebras, $\dm = \vm \semidirect \fm $, so that each element $T\in
\dm$ decomposes into a sum $T = \bov + f $ of a vector field $\bov$ and a multiplication
operator $f$. Again, 
$\fm$ defines a $\dm$-module, with $T(h) = \bov(h) + f\cdot h$ for $h\in \fm$. The Lie 
bracket or commutator is then 
$$
[T,U] = TU-UT = [\bov,\bow] + \bov(g ) - \bow(f ), \where T = \bov + f , \quad U = \bow 
+ g \in \dm.
\Eq{commutator}
$$
In local coordinates $z=\psups zm$, a \fo. differential operator has the form 
$$T = \bov + f = \sum_{i=1}^m \xi^i(z) \pdo{z^i} + f(z),\Eq{do}$$
where the coefficients $\xi ^i$ and $f$ are analytic functions of $z$. We let $\pi \colon 
\dm\to \vm $, with $\pi(\bov + f) = \bov$, denote the natural projection of a \fo. 
differential operator onto its vector field part. We can identify $\pi (T) = \bov$ with the 
``symbol'' of the differential operator $T = \bov + f$.

We are interested in studying Lie subalgebras $\g\subset \dm$, so that $\g $ is a Lie 
algebra of \fo. differential operators on the manifold $M$. Our classification
of finite-dimensional subalgebras will be 
local, leaving aside extremely interesting, but less well understood global issues.
Since all our results are local, from now on we will usually avoid explicit use
of the term ``local''.
By \eq{commutator},  the vector field part, or ``symbol'', of the algebra, defined as $\h = 
\pi
(\g) \subset
\vm$,  forms a Lie algebra of vector fields on $M$. Let $\m= \g\cap \fm $ denote the
subspace  of $\g$ consisting of all the multiplication operators in $\g$. The commutator
formula 
\eq{commutator} (with $\bow = 0$) immediately implies that $\m$ is an $\h$-module, 
meaning that if $\bov\in \h$ and $h\in \m$, then $\bov(h)\in \m$; in other words, $\m$ 
forms a representation space for the Lie algebra $\h$. 
If $T=\bov + f\in\g$, it is tempting to
define $F\colon \h\to\fm$ by $\ip F\bov=f$. However, this is incorrect, since we can
clearly add to
$T$ any function $\htilde\in\m$, and still obtain an element of $\g$ with the same symbol
$\bov$ as $T$. To remedy this, we just go to the quotient space
$\q =
\fm /\m$, which also forms an $\h$-module, and define $F:\h\to\q$ by
$$\ip F\bov=[f]\in \q,\qquad{\rm if}\quad \bov\in \h\quad{\rm and}\quad\bov+f\in\g 
.\Eq{decomp}$$
This is clearly well defined, since if $\bov\in\h$ and $T_1=\bov+f_1\in\g$,
$T_2=\bov+f_2\in\g$, then $T_1-T_2=f_1-f_2\in\g$ implies that $f_1-f_2\in\m$, and
therefore $F(\bov)=[f_1]=[f_2]$.
In view of the 
commutator formula \eq{commutator}, the fact that the differential operators in $\g$ form a 
Lie algebra implies that $F$ must satisfy the bilinear identity
$$\bov\ip F\bow - \bow \ip F\bov - \ip F{[\bov,\bow]} = 0
\quad\hbox{\rm in $\cal Q$}, \qquad \bov, \bow \in 
\h. \Eq{cocycle}$$
In the language of Lie algebra cohomology, \rf{Jacobson}, the left hand side of 
\eq{cocycle} defines the differential $\delta F$ of the linear map $F$; the fact that it 
vanishes implies that $F$ defines a 1-cocycle on $\h$ with values in the $\h$-module $\q$. 
Let $\zz$ denote the space of $\q$-valued 1-cocycles on the Lie algebra $\h$. The basic 
classification theorem, \rf{GKO}, can be stated as follows.

\Pr{lado} Let $\g\subset \dm$ be a Lie algebra of \fo. differential operators. Then 
$\g$ can be represented by a triple $(\,\h,\m,F\,)$, where: 
\ritems{$\h=\g\cap\vm $ is a Lie algebra of vector fields on $M$,\\
$\m=\g\cap \fm $ is an $\h$-module of scalar-valued functions,\\
$F\in \zz$ is a $\q = \fm/\m$--valued 1-cocycle on $\h$.} 

There are two classes of equivalence maps that preserve the basic Lie algebra 
structure of the space $\dm$. The first are the \is{changes of variables}, provided by local 
diffeomorphisms $\varphi \colon M\to M$, which act naturally on $\dm$ via
$$
\varphi _*(T) = \varphi _*(\bov + f) = d\varphi (\bov) + f \comp \varphi ^{-1} , 
\Eq{cov}
$$
where $d\varphi $ is the usual differential (push-forward) map on vector fields. 
The second are the \is{gauge transformations}, obtained by multiplying the functions in 
$\fm $ by a fixed non-vanishing function\fnote{Here, in contrast to the usual physics 
version, the gauge factor $\eta $ is not assumed to be unitary, \ie $\sigma $ is not 
necessarily purely imaginary.} $\eta (x) = e^{\sigma (x)}$. The corresponding gauge 
action on a 
differential operator is given by
$$\CG_\sigma (T) = e^{-\sigma} \cdot T\cdot e^\sigma \roq{so that} \CG_\sigma 
(\bov + f) = \bov + f + \bov(\sigma ). \Eq{gauge}$$
Thus, the effect of a gauge transformation is to modify the multiplication part of the 
differential operator by the additional term $\bov(\sigma )$.
\Df{equiv} Two Lie algebras of differential operators $\g$ and $\widetilde \g$ are 
\is{equivalent} if and only if there is an \is{equivalence map} $\Psi =(\varphi ,\CG_\sigma 
)$, consisting of a change of variables and a gauge transformation, that maps one to the 
other, so $\widetilde \g = \CG_\sigma \comp \varphi _*(\g)$.

If the operators $T$ generate a Lie algebra $\g\subset \dm$, which we identify 
with the triple $(\,\h,\m,F\,)$ as above, 
then we can interpret the gauge term $\bov(\sigma )$
in \eq{gauge} as the coboundary map $\delta \sigma \colon \h \to \q$, defined by 
$\ip{\delta \sigma }\bov = \bov(\sigma )$, associated with the 0-cochain $\sigma \in \q$ on 
the Lie algebra $\h$. 
Note that for $\sigma \in \m$ the coboundary $\delta \sigma $ 
modifies the cocycle 
$F$ in a trivial manner, since $F$ is only defined modulo elements of $\m$ anyway.  Thus, a gauge 
transformation $\CG_\sigma$ will map the Lie algebra $\g = (\h,\m,F)$ to the 
cohomologous Lie algebra $\CG_\sigma (\g) = (\h,\m,F+\delta \sigma )$. Let $\bb$ 
denote the space of $\q$-valued coboundaries, and $\hi = \zz/\bb$ the corresponding first 
Lie algebra cohomology space. We conclude that the gauge-equivalent Lie algebras 
corresponding to a given Lie algebra of vector fields $\h$ and module $\m$ are classified 
by cohomology classes $[F] \in \hi$. In other words, equivalent Lie algebras correspond to
\is{equivalent triples}
$(\h,\m,[F])$, and $(\,\widetilde \h,\widetilde \m,[\widetilde F]\,)$, meaning that
there is a change of variables $\varphi$ for which $\widetilde
\h =
\varphi _*(\h)$, $\widetilde \m = \varphi _*(\m)$, and $\widetilde F = \varphi _*\comp
F\comp \varphi_*^{-1} + \delta \sigma $.
In particular, if $F$ is cohomologous to a coboundary, then the Lie algebra $\g$ is 
equivalent,
under a gauge transformation, to the sum $\widetilde \g =
\widetilde \h + \m$ of a Lie algebra of pure vector fields and a subspace of multiplication
operators.

\Th{la} There is a one-to-one correspondence between equivalence classes of Lie 
algebras $\g$ of \fo. differential operators on M and equivalence classes of triples 
$\bigl(  \h, \m , [F] \bigr)$, where \ritems{$\h$ is a Lie algebra of vector fields,\\ $\m 
\subset \fm$ is an $\h$-module of functions,\\ $[F]$ is a cohomology class in $\hi$, 
where $\q = \fm/\m$.}

The classification of Lie algebras of \fo. differential operators therefore 
reduces to the problem of classifying triples $(\h,\m,[F])$ 
under local changes of variables. From 
the standpoint of physics, the most interesting are the finite-dimensional subalgebras
$\g\subset \dm$. In this case, most of the known results on Lie algebra cohomology,
\rf{Jacobson}, are \is{not} directly applicable since they apply to cohomology classes 
having
values in finite-dimensional modules, whereas in our case the relevant module $\q$ has 
finite
\is{co-dimension} in $\fm$. 

Based on \th{la}, there are three basic steps required to classify finite-dimensional 
Lie algebras of first order differential operators. First, one needs to classify the finite 
dimensional Lie algebras of vector fields $\h$ up to changes of variables. Secondly, for 
each of these Lie algebras, one needs to classify all possible finite-dimensional 
$\h$-modules of analytic functions. Trivial modules, valid for any Lie
algebra of vector fields, are the zero module $\m = 0$, which consists of
the zero function alone, and that containing just the constant functions,
$\m = \C$. (Indeed, for quasi-exactly solvable Lie algebras, these are the
only two admissible modules --- see below.)
Fixing the Lie algebra $\h$ and the module $\m$, the first
cohomology space $\hi$ will determine a parametrized system of inequivalent
Lie algebras of differential operators. Typically, the cohomology classes
are parametrized by one or more continuous parameters or, possibly, analytic
functions. In the case of one- or two-dimensional complex manifolds, the local
classification problem has been solved. The classification of
finite-dimensional (non-singular) Lie algebras of vector fields in one and two
complex variables appears in Lie, \rf{Liegr}, \rf{Lietrans}. The remaining 
classification of the finite-dimensional modules $\m$ and the associated
cohomology space 
$\hi$ for each of the Lie algebras of vector fields was completed by Miller, \rf{Millera}, 
and the authors, \rf{KOlado}, in one dimension, and by the authors,
\rf{GKOlado}, in two dimensions. 

\Section c Real Forms of Complex Algebras.
We now turn to the main topic of this paper, which is the extension of the known 
classification results for complex Lie algebras of differential operators to the real domain. 
We begin with a few elementary remarks concerning the complexification of real vector 
spaces. If $V$ is a real vector space, its \is{complexification} is the 
complex vector space\fnote{Tensor products are always over the real field $\R$.}
$V_\C = V\tensor \C$. Every element $u \in V_\C$ can be 
written in the form $u = v + i w$, where $v, w\in V$. Thus, $V_\C=V\oplus iV$. As 
usual, we
set $v =\Re u$, $w = \Im u$, and $\barr u = v - i w$. If $V$ is finite-dimensional, and 
$\CB =
\bsubs en$ is a basis of $V$, then it remains a (real) basis of $V_\C$. If $W\subset V$ 
is any subspace, then its complexification $W_\C = W \tensor \C$ is clearly a subspace 
of $V_\C$. We shall say that $W_\C$ is a \is{complexified subspace} of $V_\C$. Not 
every
subspace
$\What\subset V_\C$ is a complexified subspace. For example, if $V = \R^2$, so $V_\C =
\C^2$, the one-dimensional subspace $\What\subset V_\C$ spanned by the vector $(1,i)$ 
is \is{not}
a complexified subspace. The following elementary characterization of complexified
subspaces of complex vector spaces will play a crucial role in our presentation.

\Lm{complex}  Let $V_\C = V\tensor \C$ be a complexified vector space, and  let
$W_\C\subset V_\C$ be a complex subspace.  Then the following conditions are  
equivalent:
\ritems{$W_\C = W\tensor \C$ is a complexified subspace for some 
$W\subset V$.\\
$W_\C = \barr {W_\C}$,\\
 $\Re W_\C = \Im W_\C = W$ are equal, in which case $W_\C = W\tensor 
\C$.}

\Co{realbasis}  A finite-dimensional complex subspace $W_\C\subset V_\C$ of a 
complexified vector space is a complexified subspace if and only if it admits a real basis 
$\CB \subset V$, which also forms a basis for its real form $W = \Re W_\C$.  In  this
case, if $\widehat\CB = \bsubs wn$ forms an arbitrary complex basis for $W_\C$,  then a 
real
basis of $W_\C$ can be found among the elements of $\Re \widehat\CB\cup 
\Im\widehat\CB = \{\subs{\Re w}n,\subs{\Im w}n.\}$.

Clearly, the basic definitions and results stated in \section{la} can be
formulated for real manifolds without any change, other than replacing $\C$ by
$\R$ wherever it  occurs. Thus, we have the space $\fm=\fmr$ of real-valued
analytic functions defined on the real-analytic manifold $M$,
 the space $\vm$ of real-analytic vector fields on $M$, and the space
$\dm = \vm \semidirect \fm $ of analytic \fo. differential operators $T = \bov + f$.
In local coordinates $x = \psups xm$, the elements of $\dm$ take the usual form
$$T = \sum_{i=1}^m \xi^i(x) \pdo{x^i} + f(x),\Eq{rdo}$$
where $\xi ^i$ and $f$ are real-valued analytic functions on $M$. \par
In addition to 
real-valued objects, it will be convenient to consider their complex-valued counterparts,
 each of which is obtained by complexification.
 Thus we let $\fmc = \fm \tensor \C$ denote the space of
complex-valued analytic functions on $M$.  A function $f\in \fmc$ can be
decomposed into its real and imaginary parts
 $f(x) = f_1(x) + i f_2(x)$, where $f_1 = \Re f$, $f_2 = \Im f \in
\fm$. In quantum mechanics, the wave functions, while defined on a real
vector space, are generally complex-valued, and so the 
appearance here of $\fmc$ will come as no surprise. Similarly, we set $\vmc = 
\vm \tensor \C$ to be the space of complex-valued analytic vector fields on $M$. In
local coordinates, an 
element of $\vmc$ has the form 
$\bov = \sum _{i=1}^m \xi ^i(x) \partial _{x^i}$,
where 
the coefficients $\xi ^i\in \fmc$ are complex-valued analytic functions. Note that $\bov = 
\Re \bov + i\Im\bov$, where 
$$\Re \bov = \sum _{i} (\Re \xi ^i)\frac\partial{\partial x^i}, \qquad
\Im \bov = \sum _{i} (\Im \xi ^i)\frac\partial{\partial x^i}.
$$
Finally, $\dmc = \vmc \semidirect \fmc = \dm \tensor \C$ is the space of complex-valued 
analytic \fo. differential operators on $M$. Its elements have the same form as in 
\eq{rdo}, but the coefficient functions $\xi ^i$, $f$, are now allowed to be
complex-valued.

Assuming analyticity, there is an important connection
between the real and complex objects, provided by the inverse procedures of
restriction to the real domain and analytic continuation. Since we are
dealing with local issues, we can assume, for simplicity, that our differential operators are
defined on open subsets of the appropriate real or complex Euclidean space. First, suppose
$\Mhat\subset
\C^m$ is an open domain in complex
$m$-dimensional Euclidean space, and $\fhat\colon \Mhat \to \C$ a
complex-analytic function. Then the restriction of $\fhat$ to\fnote{We always
assume that $\Mhat 
\cap \R^m \ne \emptyset$.} $M = \Mhat\cap \R^m$ defines a (in general) complex-valued 
real-analytic function $f\colon M\to \C$. We let $\CR\colon \fmh \to \fmc$ denote this 
restriction map. Conversely, given a complex-valued analytic function $f\in \fmc$ defined 
on an open subset $M\subset \R^m$, its analytic continuation to the complex domain 
$\C^m$ defines a complex-analytic function $\fhat\colon \Mhat \to \C$ defined on a 
subdomain $\Mhat\subset \C^m$ such that $\Mhat\cap \R^m = M$. (Actually,
$\Mhat\cap 
\R^m$ may be strictly larger than $M$ if $f$ can be analytically continued as a real-analytic 
function.) We will always assume that, by suitably restricting the domain $\Mhat$, the 
analytic continuation $\fhat$ is a single-valued function. In our applications, the functions 
considered are, by and large, combinations of rational and exponential functions, and so 
the more technical issues associated with the process of analytic continuation do not arise. 
We let $\CC\colon \fmc \to \fmh$ denote the process of analytic continuation, so 
$\CC(f) = \fhat$. Note that the restriction and analytic continuation operators are inverses 
of each other, meaning that $\CR\comp\CC = \Id$ and $\CC\comp\CR = \Id$, provided 
$M$ and $\Mhat$ are suitably related. 

The restriction of a subspace $\mh \subset \fmh$ to the real axis will be a subspace 
$\CR(\mh)$ of the space $\fmc$ of complex-valued functions. However, in general
$\CR(\mh)$ does not arise from the complexification of a subspace of real-valued 
functions,
meaning that 
$\CR(\mh)$ need not have the form $\m \tensor \C$ where $\m\subset \fm$ is a real subspace.
\lm{complex} shows  that this happens if and only if the subspace $\CR(\mh)$ equals its 
own
complex conjugate. For example, if $\mh$ is generated by the
analytic function 
$z + i$ for $z\in \C$, then its restriction to the real axis is the one-dimensional subspace 
spanned by the complex function $x+i$, \ie is equal to the set of all functions of the 
form $c(x+i)$, where $c$ is an arbitrary complex constant. Here $\barr{\CR(\mh)}$ is
generated by $x-i$, and thus $\CR(\mh)$ is not a complexified subspace.

Similar considerations provide a correspondence between real-analytic
complex-valued vector fields and differential operators defined on $M\subset
\R^m$ and their complex-analytic counterparts, defined on $\Mhat\subset
\C^m$. The operations of restriction $\CR$ and analytic continuation $\CC$
are applied component-wise. Thus, for example, the restriction of a
complex-analytic differential operator of the form \eq{do} is obtained by
restricting its coefficients to depend on real values of $z = x+iy \in \C^m$, \ie we set $y=0$ 
in the formulae, so that
$$\CR(T) =\CR\left [ \sum_{i=1}^m \xi^i(z) \pdo{z^i} + f(z)\right ] = \sum_{i=1}^m 
\xi^i(x) \pdo{x^i} + f(x),\qquad x\in M,$$
the latter defining a complex-valued real-analytic differential operator on a 
suitable subdomain $M\subset \R^m$, \ie an element of $\dmc$. In particular, the 
restriction of
the vector field $\partial _ z$ to the real axis is merely $\partial _x$,
\is{not} $\f2(\partial _x + i \partial _y)$!
As in the case of functions, the operators $\CR$ and $\CC$ are inverses of one another 
when
acting on differential operators.

Note that the restriction and  analytic
continuation operators respect the Lie algebra structure on the space of differential 
operators; for example, the analytic continuation of the commutator bracket of two 
differential operators is the bracket of their continuations: 
$$\CC[\CD,\CE] = 
[\CC(\CD),\CC(\CE)].$$
Therefore, if $\gh \subset \dmh$ is a Lie algebra of
complex-analytic first order differential operators on $\Mhat\subset\C^m$,
its restriction to $M\subset\R^m$ is the complex Lie algebra 
$\CR(\,\gh\,)=\{\CR(T)\mid T \in\gh\}\subset \dmc$
defined on the domain $M = \Mhat\cap \R^m$.  
  The restriction $\CR(\,\gh\,)$ defines a \is{real} Lie algebra of differential operators
 if and only if it is a complexified Lie algebra, 
so that $\CR(\,\gh\,) = \gc = \g \tensor \C$, where $\g = \Re(\gc)$ is the associated
real Lie algebra. Conversely, given a 
finite-dimensional\fnote{The finite-dimensionality is important here, since otherwise
there is no guarantee that the analytic continuation of the individual 
differential operators have a common domain of definition.}
 real-analytic Lie algebra of differential operators $\g\subset \dm$, 
then there is a corresponding complex-analytic Lie algebra of differential operators $\gh = 
\CC(\gc) \subset \dmh$, defined on a suitable subdomain $\Mhat\subset \C^m$, which is 
obtained by analytically continuing the corresponding complexified Lie algebra $\gc = \g 
\tensor \C$.  Thus we deduce the fundamental correspondence between real and complex 
Lie algebras of differential operators. \par

\Pr{rcla} The analytic continuation of a real-valued finite-dimensional 
Lie algebra of real-analytic differential operators defines a finite-dimensional 
Lie algebra of complex-analytic differential operators.  
Conversely, a Lie algebra of complex-analytic differential operators determines 
a real Lie algebra of differential operators via restriction if and only if 
its restriction is a complexified Lie algebra. \par

Two complex Lie algebras of differential operators  are
\is{equivalent} if and only if there is a complex-analytic change of variables and
gauge transformation mapping one to the other. Similarly, two
real Lie algebras of differential operators are \is{equivalent} if and only if
there is a real-analytic change of variables and gauge transformation
mapping one to the other. Clearly, analytically continuing the complexifications of two
equivalent real Lie algebras produces (locally) equivalent 
complex Lie algebras. The converse,
though, is false in general: two real Lie algebras on $M$ whose complexifications have
equivalent analytic continuations are not necessarily equivalent. 

In general, by a \is{real form} of a complex Lie algebra of differential operators
 $\gt$, we mean any real Lie algebra of differential operators $\g$
 which is obtained by first applying a complex change of variables
 and gauge transformation, leading to the complex-equivalent algebra 
$\gh = \Psi (\,\gt\,)$, and then restricting to the real axis.  
\th{rcla} requires that the resulting complex-valued algebra $\gc = \CR(\,\gh\,)$
 be a complexified algebra, whereby $\g = \Re \gc$, and 
$\gc = \g \tensor \C$.  
Two different real forms of a given complex Lie algebra will always 
be analytically continuable to complex-equivalent Lie algebras, 
although the real forms may not be real-equivalent 
(even if they are isomorphic as abstract Lie algebras).  
For example, the real Lie algebras generated by 
$\dx + \dy$, $x\dx + y\dy$, $x^2\dx+y^2\dy$, 
and by $\dx$, $x\dx + y\dy$, $(x^2-y^2)\dx + 2xy\dy$ 
are both isomorphic to $\sLR2$, 
and are real forms of the complex Lie algebra with generators 
$\dz + \dw$, $z\dz + w\dw$, $z^2 \dz + w^2 \dw$.  
However, the two real forms are not equivalent --- there is no real 
change of variables mapping one to the other; see \rf{GKOreal}.
 A key problem, then, is to determine the different possible 
(real-)inequivalent real forms of a given complex Lie algebra of differential operators.\par

We now describe a general approach to this problem, based on a solution to the 
complex classification problem. Each complex Lie algebra $\gh$ of differential operators
on $\Mhat\subset\C^m$ can, in accordance with 
\th{la}, be identified with a triple $\bigl( \, \hh, \mh , [\Fhat]\, \bigr)$, where $\hh\subset
\vmh$ is a complex Lie algebra of vector fields on $\Mhat$. Clearly, if $\gh$ is to restrict 
to a
complexified Lie algebra of differential operators $\CR(\,\gh\,) = \g\tensor\C\equiv\gc$, then
$\hh$ must restrict to a complexified Lie algebra of vector fields $\CR(\hh) =\h\tensor
\C\equiv\hc
\subset\vmc$. Furthermore, the $\hh$-module 
$\mh\subset \fmh$ must restrict to a complexified module $\CR(\mh)=\mc = \m\tensor 
\C\subset \CF(M,\C)$. For example, if the complex module $\mh = \mc = \C$ consists just of 
the 
constants, then $\m = \R$ also consists of  the real constants.

Since the quotient of two complexified spaces is
complexified, the quotient space $\qc = \fmc/\mc$ is also complexified, with $\qc = \q
\tensor \C$,  where $\q = \fm/\m$. Given a complex cocycle $\Fhat\in\zh$, we define its
restriction $F_\C = \CR(\Fhat)$ to be the complex cocycle defined by
$$
\ip{F_\C}\bov = \CR\bigl(\ip \Fhat{\CC(\bov)}\bigr),
\forall\bov\in\hc.\Eq{rest}
$$
In other words,
$$F_\C = \CR\comp\Fhat\comp\CC.$$
The fact that the maps $\CR$ and $\CC$ are Lie algebra isomorphisms
implies that $F_\C$ is an element of $\zc$. In particular, the
complex vector spaces $\zh$ and $\zc$ are isomorphic:
$$
\zc = \CR\comp\zh\comp\CC.
$$
The complex conjugate
$\overline{F_\C}\colon\hc\to\barr\qc=\qc$ of a typical element $F_\C\in\zc$, defined by
$$
\ip{\overline{F_\C}}\bov = \overline{\ip{F_\C}{\barr\bov}},
\forall\bov\in\hc,\Eq{ccf}
$$
is also a cocycle by \eq{cocycle}. This shows that the space $Z^1_\C = \zc$
 of $\qc$-valued
cocycles is complexified, so $Z^1_\C = Z^1 \tensor \C$, where $Z^1 = \Re(Z^1_\C)$. 
From
\eq{ccf} it is clear that if $F_\C\in Z^1_\C$ then $\Re F_\C$ is a real $\q$-valued cocycle 
on
$\h$, and conversely any real cocycle $F\in\zr$, when extended by linearity to $\hc$, is in
$Z^1_\C$ by \eq{cocycle}. Thus, $Z^1$ can be identified with $\zr$.

Furthermore, if
$\delta\sigma$ is the coboundary of a complex-valued function $\sigma\in\qh$, then its
restriction $(\delta\sigma)_\C$ to $\zc$, as defined by \eq{rest}, is the coboundary of
$\CR(\sigma)\in\qc$, since
$$
\eeq{
\ip{(\delta\sigma)_\C}\bov = \CR(\ip{\delta\sigma}{\CC(\bov)})
 =\CR\bigl(\CC(\bov)(\sigma)\bigr)\\
 =\bov\bigl(\CR(\sigma)\bigr)=\ip{\delta\CR(\sigma)}{\bov},}
\forall\bov\in\hc.
\Eq{dalph}
$$
This means that
$$
\bc = \CR\comp\bh\comp\CC,
\Eq{brc}
$$
and the spaces $\bh$ and $\bc$ are thus isomorphic.
Given an element $\delta\beta\in\bc$, with
$\beta\in\qc$, its complex conjugate
$\barr{\delta\beta}$ is easily seen to be equal to the coboundary of $\barr\beta$. Since
$\qc$ is a complexified space, this implies that
$$\barr{\delta\beta}=\delta\barr\beta\in\delta(\,\barr\qc\,)=\delta(\qc)=\bc,$$
and therefore the space $B^1_\C = \bc$ of 1-coboundaries is also  complexified:
$$B^1_\C = B^1\tensor\C,$$
where as before it is immediate to show that $B^1 =\Re(B^1_\C)$ can be
identified with $\br$. Therefore, the resulting  cohomology space $\hic=Z^1_\C/B^1_\C 
\equiv
H^1_\C$ is isomorphic to $\hih$ and is complexified:
$$H^1_\C = \CR\comp\hih\comp\CC=H^1\tensor\C, \with H^1 = \hir.$$
This suffices to prove our basic complexification  result:
\Th{cx} Let $\hh\subset\vmh$ be a complex Lie algebra of complex-analytic vector fields, 
and let $\mh\subset\fmh$ be a complex $\hh$-module of complex-analytic functions. 
Suppose
the restrictions $\CR(\hh)\subset\vmc$ and $\CR(\mh)\subset\fmc$ are complexified 
spaces:
$$
\CR(\hh) =\h\tensor \C\equiv\hc, \qquad
\CR(\mh)=\m\tensor\C\equiv\mc.
$$
The quotient module $\qh = \fmh/\mh$ restricts to
$\CR(\qh) =\qc =\q\tensor \C= \fmc/\mc $,  with real counterpart $\q = \fm/\m$. Moreover, the
restriction $\CR\comp\hih\comp\CC$ of the
associated  cohomology space $\hih$ is also a complexified space, with
$$\CR\comp\hih\comp\CC=\hic = \hir\tensor\C\equiv H^1_\C.$$

In particular, \th{cx} implies that the real and complex cohomology spaces of a 
complex Lie algebra of vector fields and any of its real forms have the same dimension:
$$\dim_\C \hih = \dim_\C \hic =\dim_\R \hir.\Eq{dimrc}$$
Therefore, the space of complex-inequivalent Lie algebras of differential operators 
corresponding to a given $\hh\subset\vmh$ and $\mh\subset\fmh$ has the same dimension 
as
the space of real-inequivalent Lie algebras of differential operators corresponding to the
real forms $\h\subset\vm$ and 
$\m\subset\fm$. According to \co{realbasis}, if $\CB = \{[F_1],\ldots, [F_n]\}$ forms a 
basis
for  the complex cohomology space $\hic$, then a basis for the real form $\hi$ can be 
found 
among $\Re\CB \cup \Im\CB = \{[\Re F_1],\ldots, [\Re F_n],[\Im F_1],\ldots [\Im 
F_n]\}$.
In other words, exactly $n$ of the real and imaginary parts of the complex cocycles
 $\subs Fn$ will be linearly independent modulo coboundaries.  

\th{cx} implies that the problem of classifying real Lie algebras 
of differential operators can be tackled directly as follows. Let $\h$ be a real form of a 
complex Lie algebra of vector fields $\htt$. Let $\ph\colon \Mtilde\to\Mhat$ be the change 
of variables 
mapping $\htt$ to a complex-equivalent  Lie algebra 
$\hh=\varphi (\htt)$, whose restriction coincides with the complexification
of our chosen real form: $\hc = \CR(\hh) = \h\tensor \C$. If
$\mt\subset 
\fmt$ is a finite-dimensional $\htt$-module of complex-analytic functions, then
$\mh=\ph_*(\mt)$ is a \fd. $\hh$-module obtained by applying the
change of variables. We assume that its restriction $\CR(\mh)$ is a complexified
module: $\mc=\CR(\mh)=\m\tensor\C$, with $\m\subset\fm$
 a real $\h$-module. Set
$\qc=\fmc/\mc=\q\tensor\C$, with $\q=\fm/\m$. 
According to \th{cx}, the cohomology
space
$H^1_\C=\hic$ is complexified: $H^1_\C=H^1\tensor\C$, where $H^1=\hi$.
Any complex Lie algebra of differential operators with vector field part $\hh$,
represented by a cohomology class in the space $\hih\cong H^1_\C$, is
found by applying
$\varphi_*$ to a  Lie algebra of differential operators represented by an element of
$\widetilde H^1=\hit$, where 
$\qt = \fmt/\mt$. \th{cx} implies that a real basis 
for $H^1$ can be constructed by taking the real and imaginary parts of the elements of a 
basis
for $\hih$ and restricting ourselves to the real hyperplane. Moreover, \eq{dimrc} implies
that the number of
real cohomology parameters in $H^1$ will be the same as the number of complex 
cohomology
parameters in $\hih\cong H^1_\C$. In summary, we can produce a basis
for the real cohomology space $\hi$ by the following  procedure: \is i) apply the change of
variables $\varphi $ to a basis for the original  cohomology space $H^1(\htt,\qt)$, \is{ii})
restrict the resulting cohomology classes to the  real subspace, and \is{iii}) choose half of 
the
real and imaginary components of these restricted cohomology classes to form a basis.  
Note
that all the steps in this procedure are totally algorithmic, although in the third step one
should bear in mind that we are dealing with equality up to a coboundary. 
Typically, for each cocycle representing a nontrivial cohomology class in the 
complex basis, either its real part, or its imaginary part, 
provides one element of the corresponding real basis.  This procedure will
be illustrated by the examples in \section{2d}.

\Section{qes} Quasi-Exact Solvability.
A finite-dimensional Lie algebra $\g\subset \dm$ is called \is{quasi-exactly 
solvable} if it admits a non-zero finite-dimensional module (or representation space) 
$\n\subset \fm $ consisting of functions on $M$. In other words, we have $T(h)\in \n$ 
whenever $T\in \g$ and $h\in \n$. The condition of quasi-exact solvability has one 
elementary consequence that simplifies our classification procedure. We state this result 
for complex Lie algebras, although the real version is also immediate.

\Pr{qes3} If $\g$ is a quasi-exactly solvable Lie algebra represented by the triple 
$(\h,\m,[F])$ as in \th{la}, then the module $\m$ of multiplication operators is either 
trivial, $\m = 0$, or consists of constants, $\m = \C$. Moreover, if $\g = 
(\h,0,[F])$, then $\g\subset \widetilde \g = \g \oplus \C$, where $\widetilde \g= 
(\h,\C,[F])$ is a quasi-exactly solvable central extension of $\g$.

Therefore, we can assume, without loss of generality, that the only multiplication 
operators in our Lie algebra are the constant functions, \ie $\g\cap \fm = \C$. The quotient 
space $\q = \fm/\C$ is also fixed in this case.

In \rf{GKOqes}, \rf{GKO}, a complete classification of all quasi-exactly solvable 
Lie algebras of \fo. differential operators on $\C^2$ was found. A remarkable 
consequence of the general classification is that the quasi-exact solvability condition 
imposes an additional ``quantization'' condition on the cohomology parameters associated 
with the cohomology space $\hi$, meaning that these parameters must take on a discrete 
(\eg integral) set of values. (In \rf{GHKO} this result was seen, in certain cases, to be a
consequence of some fundamental algebro-geometric results governing the classification of
vector bundles over algebraic varieties.) In the present work, we shall extend the
classification to real Lie  algebras of differential operators. The basic remark is that the
quasi-exact solvability  condition is respected by the process of complexification.

\Pr{module} Let $\Mhat\subset \C^m$ be an open subset, and let $\gh\subset 
\dmh$ be a Lie algebra of complex-analytic
\fo. differential operators. Let $M=\Mhat \cap \R^m$, and 
assume that the restriction $\CR(\,\gh\,)$ is a complexified Lie algebra of 
differential operators, so that $\CR(\,\gh\,) = \g \tensor \C \equiv\gc$ for some real Lie algebra 
$\g
\subset 
\dm$. Then all three Lie algebras  $\gh$, $\gc$, $\g$ 
are quasi-exactly solvable if and only if
any one of them is quasi-exactly solvable.

\Proof
First, if $\n\subset \fm$ is a finite-dimensional module for $\g$, then its 
complexification $\nc = \n\tensor \C$ is a finite-dimensional module for $\gc$, and thus its 
analytic continuation $\nh = \CC(\nc)$ is a finite-dimensional module for $\gh$. 
Conversely, 
if $\nh$ is a finite-dimensional module for $\gh$, then its restriction $\CR(\nh)$ to $M$ 
is a finite-dimensional module for $\gc$. If $\CR(\nh) = \n\tensor \C$ is a \fd. complexified
$\gc$-module,  then it immediately follows that $\n$ is a \fd. $\g$-module . 
Otherwise, we just replace
$\CR(\nh)$ by
$\CR(\nh)+\overline{\CR(\nh)}$, which is also a finite-dimensional 
$\gc$-module and is obviously complexified.
\qed
Thus, the classification of real quasi-exactly solvable Lie algebras can be based on 
the complex classification. The only question is whether the quantization conditions 
required for quasi-exact solvability produce, upon restriction, complexified Lie algebras or 
not. This will be the case if and only if the quasi-exact solvability imposes real constraints 
on the real cohomology parameters. For example, in the planar cases appearing in the 
complex classification Tables 1, 2 and 3, the quantization conditions are all real, and hence the 
complex Lie algebra and its restriction obey the same quasi-exact solvability criteria. 
However, as we shall see, this is not the case for the other real forms of our complex Lie 
algebras, so that the real quasi-exact solvability criteria turn out to be more restrictive than 
their complex counterparts. Thus, in these cases, to determine the real quasi-exactly 
solvable Lie algebras of differential operators, we need only determine how the change of 
variables has affected the quantization condition. If the quantized values for the 
cohomology parameters remain real under the change of variables, then the real Lie algebra 
of differential operators will admit a \fd. (complex) module. If, on the other hand, the
quantized cohomology condition  is no longer real in the new variables, then the real Lie
algebra of vector fields will have no  real quasi-exactly solvable Lie algebras with nontrivial
cohomology. 

\Section{2d} The Planar Case.
Let us review what is known in the particular case of two dimensions.  In the
complex  case, there are, up to local changes of variables, a wide variety of
nonsingular Lie algebras  of vector fields, first classified by Lie,
\rf{Liegr}.  These correspond to the locally inequivalent  finite-dimensional
transformation group actions on a two-dimensional manifold; the 
nonsingularity condition avoids points where all of the vector fields
simultaneously vanish, 
\ie the transformation group has no zero-dimensional orbits.  The
classification naturally  distinguishes between the \is{imprimitive} Lie
algebras, for which there exists an invariant  foliation of the manifold, and
the \is{primitive} algebras, having no such foliation.  Since  the
construction of quasi-exactly solvable Schr\"odinger operators requires the
associated  group action to be transitive, we shall ignore the intransitive
cases here. In \rf{GKOlado},  Lie's classification was reorganized into 24
different categories, although these in turn can  be further simplified into
14 categories as given in tables 1 and 2, which are adapted from those
appearing in \rf E, and form a simplified version of the classification
appearing in our earlier papers \rf{GKOqes}, \rf{GKOlado}, 
\rf{GKO}.
  In their canonical forms, each of these Lie algebras appearing in the complex classification is a 
complexified algebra, and hence has an obvious real counterpart, obtained by restricting the 
coordinates to be real.  Moreover, according to \ths{cx}{module}, in such cases the associated 
real Lie algebras of differential operators and finite-dimensional  modules are readily 
obtained by restriction.  Interestingly, every imprimitive real Lie algebra is obtained by this 
simple procedure.  In addition to these real Lie algebras, there are precisely five additional 
primitive real Lie algebras of vector fields in two dimensions that are are not equivalent to 
any of the Lie algebras obtained by straightforward
restriction of the complex normal forms.  However, the 
complexification (or analytic continuation) of each of these five additional Lie algebras will, 
of course, be equivalent, under a complex diffeomorphism, to one of the complex normal 
forms on our list.  The complete list of these additional real forms along
with their  canonical complexifications appears in Table 4; the required
changes of variables needed to  change them into the canonical form appears
later in this section.  Therefore, to complete  the classification of all real
Lie algebras of first-order differential operators, we need only  determine the
real cohomology spaces associated with these five additional real forms, and, 
further, to determine what values of the cohomology parameters will produce
quasi-exactly  solvable algebras.  Combining these results with our earlier
complex classification will thus  produce a complete list of normal forms for
the Lie algebras of first order differential  operators  in two real variables
admitting a finite-dimensional module of smooth  functions. The full
classification is summari<ed in Tables 5--7.\par For simplicity, in the ensuing
discussion, the complex
$\h$-module will be always taken  to be $\m_C = \C$, as it will in any quasi-exactly
solvable algebra, by Proposition 8.  We will use $z,w$ for the complex
canonical coordinates, and
$x,y$ for the associated real  coordinates.\par

\Case 1 The first of the real forms is the unimodular algebra with generators
$$\qeq{\dx, \\x\dx + y\dy, \\ (x^2-y^2)\dx + 2xy\dy.}\Eq{h1}$$
The complex normal form of this Lie algebra is of Type 1.1, with generators 
$$\qeq{\dz + \dw, \\ z\dz + w\dw, \\ z^2 \dz + w^2 \dw.}\Eq{vf12}$$
Indeed, the change of variables 
$$z = x + i y,\qquad w = x - i y, \Eq{cov1}$$
maps the generators \eq{vf12} directly to \eq{h1}. The cohomology space of 
\eq{vf12} is one-dimensional, and is represented by the Lie algebra of differential 
operators spanned by
$$\qeq{\dz + \dw, \\ z\dz + w\dw, \\ z^2 \dz + w^2 \dw + c (z-w).}\Eq{do12}$$
Applying the change of variables \eq{cov1} to the associated differential operators, 
we find that the real cohomology is also one-dimensional (in accordance with 
\th{complex}) and represented by
$$\qeq{\dx, \\ x\dx + y\dy, \\ (x^2-y^2)\dx + 2xy\dy + b y,}\Eq{rdo2}$$
where the real cohomology parameter is related to the complex one via $b = 2 i c$. 
Finally, the complex quantization condition requires that $c = \f2n$, $n\in \N$, 
is a half integer.  For the real Lie algebra \eq{rdo2}, this requires that 
$b = i n$ be complex, which makes the Lie algebra no longer complexified.  
We conclude that there are no real quasi-exactly solvable Lie algebras with 
nonzero cohomology. 

Thus, the only Lie algebra of this form having a nontrivial module is the original vector 
field model \eq{h1}. In this case, the fundamental complex $\g$-modules are of the form
$$\n_m = \Span\left\{\psi^m_k \mid \ran k{2m}\right\},\Eq{h1m}$$
with $m\in \N$ a nonnegative integer,
$$\psi^m_k(x,y) = y^{m-k} R^m_k\left(x\over i\,y\right), \qquad R^m_k(t) = {d^k\over 
dt^k}\,(t^2-1)^m.\Eq{psimk}$$
It is easy to see that $\n_m$ is a complexified module for all $m\in \N$. Indeed, a 
straightforward calculation shows that the functions
$$i^k\,\psi^m_k =k!(-1)^{m-k} \sum_{{k\over2}\le j\le m}
{m\choose j}{2j\choose k} x^{2j-k} y^{m-2j},\qquad k = 0,\ldots ,2m,$$
form a real basis thereof. Every other finite-dimensional real module is a direct sum of the 
fundamental real modules $\Re \n_m$.

\Case 2 The second case to consider is the orthogonal Lie algebra generated by
$$\qeq{\bov_1 = (1+x^2-y^2)\dx + 2xy\dy, \\\bov_2 = -y\dx + x\dy, \\ \bov_3 = 
2xy\dx + (1-x^2+y^2)\dy.}\Eq{so3}$$
The group action is obtained by stereographic projection of the standard rotation 
group action on the two-sphere. The complex normal form is again of Type 1.1, as in 
\eq{vf12}, with one-dimensional cohomology, as in \eq{do12}. Under the change of 
variables 
$$z = x + i y,\qquad w = {-1 \over x - i y} = - {x+iy\over x^2 + y^2}, 
\Eq{cov2}$$
we find
$$\qeq{\dz + \dw = \f2(\bov_1 - i\bov_3), \\z\dz + w\dw = -i\bov_2, \\ z^2 \dz + 
w^2 \dw = \f2(\bov_1 + i\bov_3).}$$
Applying the change of variables \eq{cov2} to the associated differential operators 
\eq{do12}, we find that the complex cohomology is represented by a one-parameter family 
of cocycles, which we denote by $F_c$, $c\in \C$, with representative Lie algebras
$$\seq{(1+x^2-y^2)\dx + 2xy\dy + c(x+iy)\>{x^2 + y^2 + 1 \over x^2 + y^2}, \\ 
2xy\dx + (1-x^2+y^2)\dy - i c(x+iy)\>{x^2 + y^2 + 1 \over x^2 + y^2},}\qquad -y\dx + 
x\dy. \Eq{rdo3c}$$
Note that, in contrast to the first case, $F_c$ is not a real cocycle for any 
specialization of the parameter $c$. However, its real and imaginary parts, $\Re F_c$ and 
$\Im F_c$ for $c\in \R$, are themselves cocycles. Moreover, according to \th{cx}, only 
one of these two can form an independent cohomology class in $\zr$. Indeed, $\Re F_c$ 
is cohomologous to the zero cocycle since it equals the coboundary of the function
$ \f2c\log(x^2 + y^2)$. The other cocycle $\Im F_c$ therefore represents the 
complete real cohomology space which, like its complex counterpart, must be
one-dimensional. The latter cocycle can be further simplified by adding in
the coboundary of $ - c\arctan(y/x)$, leading
to the real cohomology representative
$$\qeq{(1+x^2-y^2)\dx + 2xy\dy + by, \\-y\dx + x\dy, \\ 2xy\dx + (1-x^2+y^2)\dy 
- bx,}$$
where the real cohomology parameter is related to the complex one via $b = 2 i c$. 
In the original complex coordinates, this cocycle can also be obtained by first adding in the 
coboundary corresponding to $c\log(-w)$, which effectively replaces $\dw$ by $\dw + 
c/w$, leading to the alternative complex cohomology representative
$$\qeq{\dz + \dw + \frac cw, \\ z\dz + w\dw, \\ z^2 \dz + w^2 \dw + c 
z.}\Eq{rdo3}$$
As in Case 4.1, the complex quantization condition is $c = \f2 n$, $n\in \N$, which 
requires that $b = in$. Consequently, there are no real quasi-exactly solvable Lie algebras 
with nonzero cohomology.

As before, the only Lie algebra having a nontrivial module is the original vector 
field model \eq{so3}. The fundamental $\so3$-modules are given by 
$$
\n_m = \Span\left\{\phi^m_k \mid\ran k{2m} \right\}, \qquad m\in \N, 
\Eq{mso3}
$$
where
$$
\phi^m_k(x,y) = \left( 1+x^2+y^2\over x-i\,y \right)^{m-k} R^m_k\left( 
x^2+y^2-1\over x^2+y^2+1 \right) .\Eq{phimk}
$$
Let us show, first of all, that all the modules \eq{mso3} are 
complexified modules. Indeed, from the identity 
$$
R^m_k(t) = {k!\over(2m-k)!} \>(t^2-1)^{m-k}R^m_{2m-k},
$$
\cf \rf{GKOCanada; (3.15)}, it follows that 
$$
\overline{\phi^m_k} = (-1)^{m-k} 2^{2(m-k)} {k!\over(2m-k)!}
\>\phi^m_{2m-k}. \Eq{phibar}
$$
In particular, $\phi^m_m$ is real. Therefore $\n_m$ admits the real basis 
$$\{\Re\phi^m_0,\ldots ,\Re\phi^m_{m-1},\Im\phi^m_0,\ldots
,\Im\phi^m_{m-1},\phi_m^m\}, \Eq{nbasis}
$$
and is therefore a complexified
module of dimension $2m+1$, whose real counterpart is spanned by the same
basis functions. The modules \eq{mso3} are all irreducible; indeed, it is not
hard to see that $\n_m$ is the carrier of the standard $\so3$ representation
with integer spin $m$. See \rf{GKOCanada; pp.~62--63} for the precise 
connection between the functions \eq{phimk} and the classical spherical
harmonics. 

It is also of interest to compare our results with those of
Morozov {\it et al.},
\rf{MPRST}. First, we note that the change of variables 
$$
\xi = -{2x\over x^2+y^2-1},\qquad \eta = {2y\over x^2+y^2-1}, \Eq{xieta}
$$
defines a diffeomorphism from the complement of the closed unit disc in the 
$(x,y)$ plane onto the entire $(\xi,\eta)$ plane, 
and maps the generators \eq{so3} into constant 
multiples of the differential operators 
$$
\eta\pdo\xi-\xi\pdo\eta,\quad \xi\eta\pdo\xi+(1+\eta^2)\pdo\eta,\quad 
(1+\xi^2)\pdo\xi+\xi\eta\pdo\eta, \Eq{mprst}
$$
which is precisely the basis of $\so3$ used in \rf{MPRST}. In the new $(\xi,\eta)$ 
coordinates, the basis elements \eq{phimk} of the fundamental $\so3$-module $\n_m$ are 
expressed as follows: 
$$
\phi^m_k(\xi,\eta) = (-2)^{m-k} (\xi+i\,\eta)^{k-m} \zeta^{k-m} R^m_k(\zeta),
\Eq{phimk1}
$$
where
$$
\zeta = (1+\xi^2+\eta^2)^{-1/2} = {x^2 + y^2-1\over x^2 + y^2+1}.
$$
The $\so3$-modules considered in \rf{MPRST} are $ \widetilde\n_m = \zeta ^m 
\CP\ps m$, where $\CP\ps m$ denotes the set of polynomials $P(\xi,\eta)$ of degree $\leq 
m$. However, as noted in \rf{MPRST}, these modules are certainly not irreducible. 
Indeed, it can be shown that
$$\widetilde\n_m = \DirectSum _{i=0}^{\left [\fr m2\right ]} \> \n_{m-2i},$$
where $\n_m$ is the module spanned by \eq{phimk1}, forms the decomposition of 
$\widetilde\n_m$ into irreducible summands.

\Case 3 Consider next the one-parameter family of
three-dimensional solvable Lie algebras generated by the vector fields
$$\qeq{\dx, \\ \dy,\\ \beta(x\dx + y\dy) + y\dx - x\dy,}\Eq{h3}$$
where $\beta$ is a real constant. The complex normal form has generators 
$$\qeq{\dz, \\ \dw, \\ (\beta - i)z\dz + (\beta + i)w\dw,}\Eq{vf5}$$
spanning a Lie algebra of Type
1.7 with $r=1$ and constant $\alpha = (\beta + i)/(\beta - i)$. 
Indeed, under the change of variables \eq{cov1}, we 
have
$$\ceq{\dz = \f2[\dx + i\dy], \qquad \dw = \f2[\dx -i\dy], \\ (\beta - i)z\dz + (\beta 
+ i)w\dw = \beta(x\dx + y\dy) + y\dx - x\dy.}$$
If $\alpha \notin \Q^-$, then our Lie algebra of Type 1.7 has zero cohomology; 
thus, for real $\beta$, the only case when \eq{h3} has nonzero 
cohomology is when $\beta=0$, corresponding to $\alpha = -1$. In this case, since we 
are assuming that the module is $\m = \C$, the cohomology space is one-dimensional, and 
is represented by the Lie algebra of differential operators spanned by
$$\qeq{\dz, \\ \dw + c z , \\ z\dz - w\dw.}\Eq{do5}$$
Applying the change of variables \eq{cov1}, we find that the complex cohomology 
is represented by the family of Lie algebras
$$\qeq{\dx + c x + i c y, \\ \dy + c y - i c x,\\ -y\dx + x\dy,}\Eq{rdo1a}$$
for $c\in \C$. As with Case 4.2, the real part of this cocycle forms a coboundary, 
while the imaginary part determines the real cohomology basis representative:
$$\qeq{\dx + b y, \\ \dy - b x,\\ -y\dx + x\dy.}\Eq{rdo1}$$
The real cohomology parameter is related to the complex one via $b = i c$. There 
are no complex quasi-exactly solvable algebras with nonzero cohomology in this case, and 
hence no real ones either.

For the complex Lie algebra \eq{vf5}, the most general module is a sum of the 
basic modules 
$$\widehat \n_{r,s} = \Span \set{z^jw^k}{\ran jr,\ran ks}, \qquad 
r,s\in\N.\Eq{vf5m}$$
Applying the change of variables \eq{cov1}, we see that the basic complex modules 
for the Lie algebra \eq{h3} are of the form 
$$
\n_{r,s} = \Span\set{ (x+iy)^j (x-iy)^k}{\ran jr,\ran ks}, \qquad 
r,s\in\N.\Eq{h3m}
$$
Since $\overline{\n_{r,s}}=\n_{s,r}$, the most general complexified $\g$-module 
in this case is a sum of the indecomposable modules $\n_{r,s}+\n_{s,r}$. Let $\CH\ps 
m$ denote the space of harmonic polynomials of degree at most $m$. The 
corresponding real indecomposable modules are therefore
$$\eeq{\n^\R_{r,s} = \Re(\n_{r,s}+\n_{s,r})\\ = \Span\set{ (x^2 + y^2)^j 
P_k(x,y)}{P_k \in \CH\ps k, \ran jr,\ran k{s-j}},} \qquad \ran rs.\Eq{h3mr}
$$

\Case 4 The four-dimensional real Lie algebra generated by
$$\qeq{\dx , \\\dy, \\x \dx + y\dy, \\y\dx - x\dy,} \Eq{h4}$$
has complex normal form of Type 1.9 for $r=1$, with generators 
$$\qeq{\dz,\\ \dw, \\ z\dz , \\ w\dw.}\Eq{vf6}$$
This case is completely trivial, since the complex cohomology space is
zero-dimensional, and hence there is no real cohomology either. The
fundamental complex modules are again given by \eq{h3m}, with their real
counterparts being \eq{h3mr}.

\Case 5 Finally, the six-dimensional pseudo-orthogonal Lie algebra $ \so{3,1}$ is 
realized on $\R^2$ by the generators
$${\qeq{\bov_1 = \dx, \\ \bov_2 = \dy,\\ \bov_3 = x\dx + y\dy, \\ \bov_4 = -y\dx + 
x\dy,} \atop\qeq{\bov_5 = (x^2-y^2)\dx + 2xy\dy ,\\\bov_6 = 2xy\dx + (y^2-x^2) 
\dy.}}\Eq{so31}$$
The complex normal form is of Type 1.4, with generators 
$$\qeq{\dz,\\ \dw, \\ z\dz,\\ w\dw, \\ z^2 \dz,\\ w^2 \dw,}\Eq{vf11}$$
spanning the Lie algebra $\so{4,\C}\cong \sL2\oplus\sL2$.
The complex cohomology space for \eq{vf11} is two-dimensional, and is represented by
$$\qeq{\dz,\\ \dw, \\ z\dz,\\ w\dw, \\ z^2 \dz + c_1 z,\\ w^2 \dw + c_2 
w.}\Eq{do11}$$
Under the change of variables \eq{cov1}, we note that
$$\qeq{ z^2 \dz = \bov_5 + i \bov_6, \\ w^2 \dw = \bov_5 - i \bov_6.}$$
Therefore, the real cohomology is also two-dimensional, and represented by
$${\qeq{\dx, \\ \dy,\\ x\dx + y\dy, \\ -y\dx + x\dy, }\atop \qeq{(x^2-y^2)\dx + 
2xy\dy + b_1x + b_2 y,\\2xy\dx + (y^2-x^2) \dy - b_2x + b_1 y.}}$$
The real cohomology parameters are related to the complex ones via 
$$b_1 = c_1 + c_2,\qquad b_2 = i(c_1 - c_2).$$
Finally, the complex quantization conditions $c_1 = -n_1$, $c_2 = -n_2$, $n_1, 
n_2\in \N$, require that the real parameters assume the values $b_1 = -2n$, $b_2 = 0$, 
where $n = n_1 = n_2$. Therefore, this real Lie algebra of vector fields determines a 
nontrivial family of quasi-exactly solvable Lie algebras of differential operators, 
parametrized by the integral cohomology parameter $n$. This finishes the proof of the 
following theorem.

\Th{realqes} Among the five additional real Lie algebras of planar vector fields in $\R^2$, 
the 
only one admitting a nonzero real-valued quasi-exactly solvable cohomology class is \ro (a 
central extension of\/\ro ) $\so{3,1}$, 
for which the corresponding Lie algebra of first-order 
differential operators is spanned by
$${ \qeq{T^0 = 1,\\ T^1 = \dx,\\ T^2 = \dy,\\ T^3 = x\dx+y\dy,\\ T^4 = y\dx-x\dy,}\atop
\qeq{ T^5 = (x^2-y^2)\dx+2xy\dy-2nx,\\ T^6 = 2xy\dx+(y^2-x^2)\dy-2ny,}}\Eq{so31n}$$
where $n$ is a non-negative integer.

In this case, the associated complex module is the polynomial module $\n_{n,n}$, as 
defined by \eq{h3m}; see also \eq{h3mr} for its real counterpart.
Finally, we note the remarkable fact that all the preceding cases form subalgebras of 
the latter case.

\Th{sub} Every finite-dimensional Lie algebra of first order differential operators with
vector field part given by one of the five additional primitive real forms listed in
Table 4 is a subalgebra of one of the pseudo-orthogonal Lie algebras given by
\eq{so31n}.\par

\Section{do} New Quasi-Exactly Solvable $\so{3,1}$ Potentials.

Now that we have completed the classification of real quasi-exactly solvable 
Lie algebras of differential operators, we can apply the additional canonical forms 
to construct new quasi-exactly solvable quantum Hamiltonians in two variables. 
We begin by recalling the basic connection between Lie algebras of first-order differential
operators and quasi-exactly solvable quantum systems. Let 
$M$ be a real or complex manifold.
The algebra $\CD^*(M)$ consisting of all differential 
operators
on $M$ can be identified with the universal enveloping algebra of the space 
$\dm$ of first-order differential operators. 
There is a natural filtration of $\CD^*(M)$ by the
subspaces $\CD^n(M)$ consisting of all differential operators of order at most $n$;
equivalently, we may define $\CD^n(M)$ as the space of all polynomials
 of degree at most $n$ in the
\fo. differential operators. In local coordinates, a typical element of 
$\CD^n(M)$ can be written in the form
$$L = \sum _{0\leq j_1+\cdots+j_m\leq n} b_{j_1\dots j_m}(z) \; \frac{\partial
^{j_1+\cdots+j_m}}{(\partial z^1)^{j_1}\dots (\partial z^m)^{j_m}}.
\Eq{diffop}$$ 
In physical applications, the differential operator is required to be Hermitian and real.
\Df{laop} A differential operator is called \is{Lie-algebraic} 
if it lies in the universal enveloping algebra of a 
finite-dimensional subalgebra $\g\subset \dm$ \par
In other words, any Lie-algebraic operator $L$ can be written as a polynomial 
(with constant coefficients) in the generators $T^a$, 
which are first order differential operators, of the Lie algebra $\g$.  
\Df{qesop} A differential operator is called \is{quasi-exactly solvable} if it is Lie-
algebraic for a quasi-exactly solvable Lie algebra $\g\subset \dm$. \par
  In this case, the finite-dimensional module $\n\subset \fm$ associated to
the Lie algebra $\g$ is invariant under  the quasi-exactly solvable operator
$L$, so that
$L(\n)\subset \n$, and hence $L$ restricts  to define a finite-dimensional
matrix operator on $\n$, whose eigenvalues will (provided the  functions in
$\n$ are normalizable) form the algebraic part of the spectrum of $L$.   \par

We are interested in constructing examples of quasi-exactly solvable 
Schr\"odinger operators.  By a
 \is{Schr\"odinger operator} we mean a 
second order differential operator  of 
the form
$$ H = -\Delta + V(x); \Eq{so}$$
here
$$\Delta=\sum_{i,j=1}^m g^{ij}(x^1,\dots,x^m)\nabla_i\,\nabla_j,$$
is the Laplacian (kinetic energy) operator on the \fd. real Riemannian manifold
$M$ with contravariant metric $(g^{ij})$, 
and $\nabla_i$ is the covariant derivative associated
to this metric.  (The physical units are taken so that $\hbarq=2m=1$.)  
Let $\g$ be one of our real
 quasi-exactly solvable Lie algebras of first order differential operators 
written in canonical form.  The most general (real) second-order differential operator $L$ 
which is quasi-exactly solvable with hidden symmetry algebra
$\g$ takes the form
$$
L = \sum_{a,b=1}^r C_{ab}{T^a}{T^b} + \sum_{a=1}^r C_{a}T^{a} + C_0,
\Eq t
$$
where the $T^a$ are the generators of $\g$, and $C_{ab}$, $C_a$ are real
constants.  Now, in general $L$ need not be a Schr\"odinger operator  \eq{so}; however,
$L$ can sometimes be transformed into a \sc. operator by applying a suitable change of
variables $\ph$ and gauge transformation $\CG_\sigma$. The transformed operator
$L=e^{-\sigma}\comp
\ph_*(L)\comp e^\sigma$ is then a \qes. operator with respect to the transformed algebra
$\CG_\sigma\comp\ph_*(\g)\cong\g$,
with generators $\Ttilde^a = \CG_\sigma\comp\ph_*(T^a)$, 
and so, under the extra assumption that the transformed wave
functions are normalizable, a finite sector of its spectrum can be 
algebraically computed.  \par
We recall now the necessary and sufficient 
conditions
under which the operator \eq t can be transformed into a \sc. operator $H$, \cf for
example \rf{GKO2}. First of all, we rewrite $L$ in the form
$$
L = -\sum_{i,j=1}^m g^{ij}(x) \,\frac{\partial^2}{\partial x^i\,\partial x^j}+
\sum_{i=1}^m b^i(x) \frac\partial{\partial x^i}+c(x).
$$
We must first require that $L$ be \is{elliptic}, meaning that the 
quadratic form associated to the symmetric matrix $(g^{ij}(x))$ be
positive definite everywhere. We may thus 
interpret
the functions $g^{ij}(x)$ as the contravariant components of a Riemannian metric
$$
ds^2 = \sum_{i,j=1}^m g_{ij}(x) \,dx^i\,dx^j,\Eq{metric}
$$
where $(g_{ij}(x))$ is the inverse of  $g^{ij}(x)$. It is convenient to express $L$ in 
covariant
form as follows:
$$
L = -\sum_{i,j=1}^m g^{ij}  (\nabla_i-A_i)(\nabla_j-A_j)+V,
$$
where $\nabla_i$ is the covariant derivative associated to the metric \eq{metric}, and
$$
\eeq{
A_j = \sum_{j=1}^m g_{ij} \left[\frac{b^i}2+
\frac1{2\,\sqrt g}\sum_{k=1}^m\frac\partial{\partial x^k}(\sqrt g g^{ik})\right],&
A^i = \sum_{j=1}^m g^{ij} A_j,\\
V = c+\sum_{i=1}^m \left[A_i\,A^i -
\frac1{\sqrt g}\frac\partial{\partial x^i}(\sqrt g A^i)\right],}
$$
with
$g = \det(g_{ij})$.  We define the \is{magnetic one form} associated with such an operator
to be 
$$
\omega = \sum_{i=1}^m A_i\,dx^i.
$$

\Th{soeq} The necessary and sufficient condition for an elliptic
second order differential operator $L$ to be equivalent to a Schr\"odinger operator
is that its magnetic one form be closed: 
$$
d\omega =0.  \Eq{clc}
$$

For a given Lie algebra of \do.s $\g$, \eqE{clc} is equivalent to a set of algebraic equations 
in
the coefficients $C_{ab}$ and $C_a$, which are called the
\is{closure conditions}.  In the complex case, these conditions were extensively analyzed
in \rf{GKO2}, although their complete solution, and hence the complete
classification of quasi-exactly solvable Schr\"odinger operators, remains problematic.

We now proceed to construct a few new examples of \qes. Hamiltonians based on the
classification of the real Lie algebras of first-order \do.s developed in the previous section.
According to \th{sub}, all the additional real canonical forms are 
subalgebras of Case 4.5, so that there is no loss of generality in working 
exclusively
with $\so{3,1}$. Interestingly, although some of the additional real normal forms found in 
the
previous section have already been used to obtain \qes. potentials 
--- see, for instance,
\rf{MPRST},
\rf{Shifman},
\rf{ShifTurb} for $\so3$ --- we are not aware of any \qes. potential that has been {\it 
explicitly}
linked to $\so{3,1}$ in the literature. However, the algebra $\so{3,1}$ 
does occur in the chain of subalgebras leading to the $\so{4,2}$ 
spectrum generating algebra for the hydrogen atom, arising from the
conserved angular momentum and Runge--Lenz vectors;
see the review article \rf{ACP}.
The only example of \qes. $\so{3,1}$ potential that we
know of is the remarkable multiparameter family recently constructed by
Zaslavskii, \rf{Zas; eq.~(31--33)}. 
Although the Hamiltonians in this family were obtained without
explicitly using Lie-algebraic techniques, it can be shown (\cf \rf{Zas2}) that they arise 
from
Hamiltonians that are \qes. with respect to the Lie algebra $\sL2\oplus\sL2$ of Type 1.4,
 when one performs the change of variables \eq{cov1},
 and requires that the resulting second-order 
differential operator have real coefficients. 
In fact, these Hamiltonians could have been
obtained much more directly by starting with the most general Lie-algebraic
second-order \do.  by taking an arbitrary
polynomial of degree two in the generators of
$\so{3,1}$, and imposing that 
\is i) the coordinates for the induced metric be isothermal, 
and
\is {ii}) the closure conditions be satisfied. It can be shown that the most
general Hamiltonian satisfying these two conditions depends on 15 real parameters 
satisfying 9
algebraic constraints. Thus, the set of all such Hamiltonians is parametrized by
 a 6-dimensional
algebraic variety. Although the number of essential parameters in Zaslavskii's 
multiparameter
family is 6, we shall now show that the latter family is only one 
of several components of the
variety. Indeed, we now present a different 6-parameter family of Hamiltonians satisfying 
the
above two conditions.

Indeed,  consider the family of second-order \do.s \eq t defined by the following choice of 
the
coefficients $C_{ab}$ and $C_a$ :
$$
\qeq{
(C_{ab}) =
\pmatrix{ \alpha  & 0 & 0 & 0 & \mu / 2 & -\beta  \cr 0 & \alpha  & 
  0 & 0 & 0 & 0 \cr 0 & 0 & 0 & \beta  & \gamma  & \lambda  \cr 0 & 0 & 
  \beta  & \mu  & -\lambda  & \gamma  \cr \mu / 2 & 0 & \gamma  & 
  -\lambda  & \nu  & 0 \cr -\beta  & 0 & \lambda  & \gamma  & 0 & \nu  \cr 
   },\\
(C_a) = -2\,n\bigl(0,0,0,\beta,\gamma,\lambda\bigr),}
$$
where $T^1,\ldots, T^6$ are the $\so{3,1}$ generators given by \eq{so31n}.
(We omit $T^0=1$ without loss of generality.)
A long but straightforward calculation shows that the closure conditions are satisfied. The
associated metric is
$$
ds^2 = A^{-1}\,(dx^2+dy^2),
\Eq{exm}
$$
with
$$
A = \alpha  + \mu \,x^2 - 2\,\beta \,x\,y + 
  2\,\gamma \,x\,( x^2 + y^2)  + 
  2\,\lambda \,y\,( x^2 + y^2 )  + 
  \nu \,( x^2 + y^2 ) ^2,
$$
so that the $(x,y)$ coordinates are isothermal. The Gaussian curvature is given by
$$
\kappa =
A^{-1}\bigl[\Strut
                a+b\,x+c\,y+d\,(x^2+y^2)+e\,xy+f\,x^2+g\,x\,(x^2+y^2)+h\,y\,(x^2+y^2)
                +k\,(x^2+y^2)^2
                \bigr],
$$
with
$$
\Eqalign{%
a &= \alpha \,\mu,&
b &= 8\,\alpha \,\gamma,&
c &= 8\,\alpha \,\lambda ,\cr
d &= 2\,(4\,\alpha \,\nu -\beta ^2 ),&
e &= 2\,\beta \,\mu,&
f &= -\mu ^2,\cr
g &= 2\,( 2\,\beta \,\lambda  - \gamma \,\mu ),&
h &= 2\,( 2\,\beta \,\gamma  + \lambda \,\mu),&
k &= \mu \,\nu-2\,\gamma ^2 - 2\,\lambda ^2 .\cr
}
$$
Since the closure conditions are satisfied, we know that $L$ is equivalent under a gauge
transformation to a \sc. operator \eq{so} on the open subset of $\R^2$ where $A$ is 
positive,
with metric given by \eq{exm}. In fact, we have
$$
H = \eta\,T\,\eta^{-1} = -\Delta+V(x,y), \where
\eta = A^{-n/2},
$$
and the potential is
$$
\eqalign{%
V &= {n\,\left( n + 2 \right)\over A}
\left[\vphantom{{T^2}^T}
     -4\, \alpha\,(\gamma\,x+\lambda \,y)
     + \mu^2\,x^2
     - 2\, \beta \,\mu \,x\,y
     +( \beta ^2 - 4\,\alpha \,\nu  ) \,( x^2 + y^2 )\right.\cr
     &\hskip3em\left.{}+ \vphantom{{T^2}^T}2\,x\,(\gamma \,\mu -\beta \,\lambda)\,( x^2 
+ y^2)
        - 2\,\beta \,\gamma \,y\,(x^2+y^2) 
     + ( \gamma ^2 + \lambda ^2 ) \,
        ( x^2 + y^2 ) ^2        
     \right].\cr}
$$
It should be noted that for generic values of the parameters neither the above solution of the
closure conditions (nor the one given by Zaslavskii), satisfy the additional condition that 
the
associated metric be positive definite in all of $\R^2$.

One can obtain many other multiparameter families of $\so{3,1}$ potentials, for
instance by dropping the condition that the $(x,y)$ coordinates be isothermal for the metric. 
We
shall content ourselves with the following example, in which
$$
\qeq{
(C_{ab}) = \diag(\alpha,\alpha,\beta,\gamma,\lambda,\lambda),\\
(C_a) = 0.}
$$
Again, the closure conditions are satisfied by the above choice of coefficients. The 
associated
contravariant metric tensor $(g^{ij})$ is given by
$$
\eeq{
g^{11} =\alpha  + \beta \,x^2 + \gamma \,y^2 + 
              \lambda \,( x^2 + y^2)^2,\\
g^{12} = ( \beta  - \gamma ) \,x\,y,\\
g^{22} =\alpha  + \gamma \,x^2 + \beta \,y^2 + 
 \lambda \,(x^2 + y^2)^2.
}
$$
The Gaussian curvature is
$$
\kappa = {(-\beta  + 3\gamma  ) (\alpha ^2 + \lambda ^2 r^8) +  
    2( \beta \gamma  + 4\alpha \lambda  ) r^2 (\alpha +\lambda r^4)
  +  2\alpha  \lambda( 5\beta  + \gamma  ) r^4
\over ( \alpha  + \gamma \,r^2 + \lambda \,r^4 )^{2}}
$$
with
$r^2 = x^2+y^2$.  
In contrast with the previous case, if the parameters $\alpha$, $\beta$, $\gamma$ and
$\lambda$ are positive, then the metric is non-singular and positive definite for $(x,y)$
ranging over all of $\R^2$.

As before, the fact that the closure conditions are satisfied guarantees
the existence of a gauge factor $\eta$ such that $H=\eta\,L\,\eta^{-1}$ is a
\sc. operator. If
$$
\rho =   4\,\alpha \,\lambda - \beta ^2,
$$
the gauge factor is given by
\par
$$
\eta = \cases{\openup6\jot
\exp\left(\displaystyle \frac{n\,\beta}{\sqrt{\rho}}
\,\arctan\displaystyle \frac{\beta  + 2\,\lambda \,r^2}{\sqrt{\rho}}\right)\,
  \left( \alpha  + \beta \,r^2 + \lambda \,r^4 \right)^
    {-{1\over 4} - {n\over 2}}\,
  \left( \alpha  + \gamma \,r^2 + \lambda \,r^4 \right)^
    {1\over 4},
    &$\; \rho>0$;\cr
 \exp\left(\displaystyle -\frac{2\,n\,\alpha}{2\,\alpha  + \beta \,r^2}\right)\,
\left( 2\,\alpha  + \beta \,r^2 \right) ^{-{1\over 2} - n}\,
     \left( 4\,\alpha ^2 + 4\,\alpha \,\gamma \,r^2 + 
          \beta ^2\,r^4 \right)^{1\over 4},
          &$\; \rho=0$;\cr
     \left( \alpha  + \beta \,r^2 + \lambda \,r^4 \right)^
       {-{3\over 4} - n}\,
       \left( \alpha  + \gamma \,r^2 + 
          \lambda \,r^4 \right)^{1\over 4}\,
     \left(
   \displaystyle \frac{2\,\lambda \,r^2+\beta-\sqrt{-\rho }}
            {2\,\lambda \,r^2+\beta+\sqrt{-\rho}}\right)^{\frac{n\,\beta }{\sqrt{-\rho}}},
        &$\; \rho<0$.}
$$
\par\noindent
In all cases, the expression for the potential $V$ is
$$
\eqalign{4 V&=
\frac{16\,\alpha \,\beta \,n\,( 1 + n )  + 
      r^2\,\left[ \beta ^2\,( 3 + 16\,n + 16\,n^2) 
      -4\,\alpha \,\lambda \,( 3 + 8\,n + 4\,n^2 )\right]}{
         \alpha  + \beta \,r^2+ \lambda \,r^4}\cr
  &\qquad {}+\frac{5\,( \beta  - \gamma ) \,
       ( 4\,\alpha \,\lambda - \gamma ^2)  + 
      3\,\lambda \,( 2\,\beta \,\gamma  - 3\,\gamma ^2+ 
         4\,\alpha \,\lambda  ) \,r^2}{
    \lambda \,( \alpha  + \gamma \,r^2 + \lambda \,r^4)}\cr
  &\qquad {}-\frac{5\,( \beta  - \gamma) \,
      (4\,\alpha \,\lambda  -\gamma^2) \,
      ( \alpha  + \gamma \,r^2) }{ 
    \lambda \,( \alpha  + \gamma \,r^2 + \lambda \,r^4)^2}
    +4\gamma  - 4\beta \,( 1 + 2\,n ) ^2,\cr
    }
$$
with $r^2=x^2+y^2$.
Since the potential is a function of $r$ only, it is natural to look for eigenfunctions of
$H$ which depend on $r$ only. When this is done, it can be shown that one ends up with 
an
effective Hamiltonian on the line which is an element of the enveloping algebra of the
standard realization of $\sL{2,\R}$ in one dimension. Thus, no new \qes. one-dimensional
potentials are obtained by reduction of the above \qes. $\so{3,1}$ potential. 
This lends additional support to the observation of \rf{GKO2}
 that reduction of two-dimensional 
quasi-exactly solvable Schr\"odinger operators does not lead to any new
one-dimensional quasi-exactly solvable operators.  However, a full explanation of
this fact remains obscure.

\medbreak\indent{\it Acknowledgment:} It is a pleasure to thank the referees for
useful comments.

\page
\References
\page
\input rf.tab

\ends
\bye